\documentstyle[12pt,epsfig]{article}
\begin{document}

\begin{titlepage}
\rightline{April 2010}
\vskip 2cm
\centerline{\Large \bf
A CoGeNT confirmation of the DAMA signal}

\vskip 2.2cm
\centerline{R. Foot\footnote{
E-mail address: rfoot@unimelb.edu.au}}

\vskip 0.7cm
\centerline{\it School of Physics,}
\centerline{\it University of Melbourne,}
\centerline{\it Victoria 3010 Australia}
\vskip 2cm
\noindent
The CoGeNT collaboration has recently
reported a rising low energy spectrum in their ultra low noise Germanium detector. 
This is particularly interesting as the energy range probed by CoGeNT overlaps with the
energy region in which DAMA has observed their annual modulation signal.
We show that
the mirror dark matter candidate can simultaneously explain both
the DAMA annual modulation signal and the rising low energy spectrum observed by CoGeNT.
This constitutes a model dependent confirmation of the DAMA signal and adds weight to the
mirror dark matter paradigm.

\end{titlepage}

The CoGeNT experiment operating in the Soudan Underground Laboratory has recently 
presented new results in their search for light mass dark matter interactions\cite{cogent}.
With a low energy threshold of $0.4$ keVee, they have observed a rising low energy
spectrum which is not readily explainable in terms of known background sources. 
The energy region probed by CoGeNT overlaps with the energy region
in which the DAMA experiments\cite{dama1,dama2} have observed 
their impressive annual modulation signal, and thus it is natural to interpret
the CoGeNT excess in terms of dark matter interactions.
The purpose of this paper is to examine the compatibility of the CoGeNT and DAMA experiments
utilizing the mirror dark matter framework which has proven successful in
explaining the direct detection experiments\cite{mm,mm2,mm3}.

Recall, mirror dark matter posits that the inferred dark matter in the Universe arises from
a hidden sector which is an exact copy of the standard model sector\cite{flv} (for a review
and more complete list of references see ref.\cite{review})\footnote{
Note that successful big bang nucleosynthesis and successful
large scale structure requires effectively asymmetric initial
conditions in the early Universe, $T' \ll T$ and $n_{b'}/n_b \approx 5$. 
See ref.\cite{some} for further discussions.}.
That is, 
a spectrum of dark matter particles of known masses are predicted: $e', H', He', O', Fe',...$ (with
$m_{e'} = m_e, m_{H'} = m_H,$ etc). 
The galactic halo is then presumed to be composed predominately of a spherically distributed 
self interacting mirror particle plasma comprising these particles\cite{sph}. 

In addition to gravity, ordinary and mirror particles
interact with each other via (renormalizable) 
photon-mirror photon kinetic mixing\cite{flv,he}:
\begin{eqnarray}
{\cal L}_{mix} = \frac{\epsilon}{2} F^{\mu \nu} F'_{\mu \nu}
\end{eqnarray}
where $F_{\mu \nu}$ ($F'_{\mu \nu}$) is the ordinary (mirror) $U(1)$ gauge boson 
field strength tensor. This interaction enables mirror charged particles to couple to
ordinary photons with electric charge $q = \epsilon e$.

A dissipative dark matter candidate like mirror matter can only survive
in an extended spherical distribution in galaxies without collapsing
if there is a substantial heating mechanism to replace the energy lost
due to radiative cooling. In fact, ordinary supernova explosions
can plausibly supply the required heating if the photon and mirror
photon
are kinetically mixed with $\epsilon \sim 10^{-9}$\cite{sph}.
For kinetic mixing of this magnitude 
about half of the total energy emitted in ordinary Type II Supernova
explosions ($\sim 3\times 10^{53}$ erg) will be in the form of light
mirror particles ($\nu'_{e,\mu,\tau}$, $e'^{\pm}, \gamma'$)
originating from kinetic mixing induced plasmon decay into $e'^+ e'^-$
in the 
supernova core\cite{rafelt}. It turns out that this energy source matches (within uncertainties) 
the energy lost from the galactic halo due to radiative cooling and is plausibly the
mechanism which stabilizes the halo from collapse\cite{sph}.

Mirror dark matter explains the DAMA annual modulation signal via kinetic mixing induced
elastic (Rutherford) scattering of the dominant mirror metal component, $A'$, off target nuclei.
[The $He'$ and $H'$ components are too light to give a signal above the DAMA energy threshold]. 
Such elastic scattering can explain the normalization and energy dependence of the DAMA annual 
modulation amplitude consistently with the null results of other experiments, 
and yields a measurement of $\epsilon \sqrt{\xi_{A'}}$ and $m_{A'}$\cite{mm,foottoappear}:
\begin{eqnarray}
\epsilon \sqrt{\xi_{A'}} &\approx & (7 \pm 4)\times 10^{-10}, \nonumber \\
\frac{m_{A'}}{m_p} &\approx & 22\pm 8
\label{bla}
\end{eqnarray}
where $\xi_{A'} \equiv n_{A'}m_{A'}/(0.3 \ GeV/cm^3)$ is the halo mass fraction of the species
$A'$, $m_p$ is the proton mass. 
The measured value of $m_{A'}/m_p$ is consistent with $A' \sim O'$, which
by analogy with the ordinary matter sector would be the naive expectation,
while kinetic mixing with magnitude $\epsilon \sim 10^{-9}$ is consistent with the value
required to stabilize the halo from collapse and also laboratory, 
astrophysical and cosmological constraints\cite{lab1}.

%

The event rate in an experiment like CoGeNT depends on the interaction cross 
section and halo distribution function.
The cross section for a mirror nucleus (with mass and atomic numbers $A',\ Z'$)
to elastically scatter off an ordinary nucleus (presumed at rest with mass 
and atomic numbers $A,\ Z$)  is given by\cite{mm}:
\begin{eqnarray}
{d\sigma \over dE_R} = {\lambda \over E_R^2 v^2}
\label{cs}
\end{eqnarray}
where 
\begin{eqnarray}
\lambda \equiv {2\pi \epsilon^2 Z^2 Z'^2 \alpha^2 \over m_A} F^2_A (qr_A) F^2_{A'} (qr_{A'}) \
\end{eqnarray}
and $F_X (qr_X)$ ($X = A, A'$) are the form factors which
take into account the finite size of the nuclei and mirror nuclei.
A simple analytic expression for
the form factor, which we adopt in our numerical work, is the one
given by Helm\cite{helm,smith}.

The halo mirror particles are presumed to form a self interacting plasma at an isothermal temperature $T$.
This means that the halo distribution function is given by a Maxwellian distribution,
\begin{eqnarray}
f_i (v) &=& e^{-\frac{1}{2} m_i v^2/T}  \nonumber \\  
 &=& e^{-v^2/v_0^2[i]} 
\end{eqnarray}
where the index $i$  labels the particle type [$i=e', H', He', O', Fe'...$].
The dynamics of such a mirror particle plasma has been investigated previously\cite{sph,mm},
where it was found that the condition of hydrostatic equilibrium implied that the
temperature of the plasma satisfied:
\begin{eqnarray}
T \simeq  {1 \over 2} \bar m v_{rot}^2 \ ,
\label{4}
\end{eqnarray}
where $\bar m = \sum n_i m_i/\sum n_i$ is the mean mass of  
the particles in the plasma, and $v_{rot} \approx 254$ km/s is the galactic rotational velocity\cite{reid}.
Assuming the plasma is dominated by $e', H', He'$ and is completely ionized, 
a reasonable approximation since it turns out that
the temperature of the plasma is $\approx \frac{1}{2}$ keV, then:
\begin{eqnarray}
{\bar m \over m_p} = {1 \over 2 - \frac{5}{4} Y_{He'}} 
\end{eqnarray}
where $Y_{He'}$ is the $He'$ mass fraction. Mirror BBN studies\cite{bbn} indicate that $Y_{He'} \simeq 0.9$, which
is the value we adopt in our numerical work.
Evidently, the velocity dispersion of the particles in the mirror matter halo depends
on the particular particle species and satisfies:
\begin{eqnarray}
v_0^2 [i] = v_{rot}^2 \frac{\overline{m}}{m_i} \ .
\end{eqnarray}
Thus for the mirror metal component, $A'$, we expect a very narrow velocity 
dispersion, $v_0 [A'] \ll v_{rot}$,
which is in contrast with non self interacting WIMPs which feature $v_0 \simeq v_{rot}$.

The CoGeNT experiment measures the absolute rate with a Germanium target, and the
recoil energy range covered by CoGeNT overlaps with the DAMA energy range. This means that the
CoGeNT experiment
should be sensitive to the same dark matter component as DAMA, $A'$.
The event rate is given by:
\begin{eqnarray}
{dR \over dE_R} = N_T n_{A'} 
\int^{\infty}_{|{\textbf{v}}| > v_{min}}
{d\sigma \over dE_R}
{f_{A'}({\textbf{v}},{\textbf{v}}_E) \over k} |{\textbf{v}}|
d^3v 
\label{55}
\end{eqnarray}
where $N_T$ is the number of target nuclei per kg of detector and  
$n_{A'} = \rho_{dm} \xi_{A'}/m_{A'}$ is the number density of halo mirror nuclei $A'$ at the Earth's
location (we take $\rho_{dm} = 0.3 \  GeV/cm^3$).
Here ${\bf{v}}$ is the velocity of the halo particles relative to the
Earth and ${\bf{v}}_E$ is the
velocity of the Earth relative to the galactic halo.
The integration limit, $v_{min}$, is given by the kinematic relation:
\begin{eqnarray}
v_{min} &=& \sqrt{ {(m_{Ge} + m_{A'})^2 E_R \over 2 m_{Ge} m^2_{A'} }}\ .
\label{v}
\end{eqnarray}
The halo distribution function is given by, 
$f_{A'} ({\bf{v}},{\bf{v}}_E)/k = (\pi v_0^2[A'])^{-3/2} exp(-({\bf{v}}
+ {\bf{v}}_E)^2/v_0^2[A'])$. The integral, Eq.(\ref{55}), can easily be evaluated in terms
of error functions\cite{mm,smith} and numerically solved.

To compare with the measured event rate, we include detector resolution effects 
and overall detection efficiency:
\begin{eqnarray}
{dR \over dE_R^m} = \epsilon_f (E_R^m) {1 \over \sqrt{2\pi}\sigma_{res} } 
\int {dR \over dE_R} e^{(E_R - E_R^m)^2/2\sigma^2_{res}} dE_R 
\end{eqnarray}
where $E_R^m$ is the measured energy and $\sigma^2_{res} = \sigma_n^2 + (2.35)^2 E_R \eta F$
with $\sigma_n = 69.4$ eV, $\eta = 2.96$ eV and $F =0.29$\cite{cogent,cog2}.
The detection efficiency, $\epsilon_f (E_R^m)$, was given in figure 3 of ref.\cite{cogent},
which we approximate via
\begin{eqnarray}
\epsilon_f (E_R^m) \simeq {0.87 \over 1 + (0.4/E_R^m)^{6} }.
\end{eqnarray}
The energy is in keVee units (ionization energy). For nuclear recoils
in the absence of any channeling, $keVee = keV/q_{Ge}$, where $q_{Ge} \simeq 0.21$ is the 
relevant quenching factor for Ge at low energies\cite{collarnew}. Channeled events, where target atoms travel down crystal
axis and planes, have $q_{Ge} \simeq 1$.

Recently it has been found\cite{newstudy} that the channeling fraction is likely to be very small
($< 1\%$) in the energy range of interest in contrast with the previous study by the DAMA Collaboration\cite{chan}. 
It is argued\cite{newstudy} that the previous analysis
did not take into account the fact that the scattered atoms originate from lattice sites 
and hence cannot be easily channeled.
In view of these developments, we shall assume that 
the channeling fraction is indeed negligible.

We fit the CoGeNT data in the low recoil energy range assuming $A'$
dark matter and that the background is an energy independent constant,
together with two Gaussians
to account for the $^{65}Zn$ (1.1 keV) and $^{68}Ge$ (1.29 keV) 
L-shell electron capture lines.
Fixing $m_{A'}/m_p = 20$ and $v_{rot} = 254$ km/s, as an example, we
find a best fit
of $\chi^2_{min} = 14.8$ for $21-4 = 17$ degrees of freedom, with
$\epsilon \sqrt{\xi_{A'}} = 6.2\times 10^{-10}$.
This fit of the CoGeNT data is shown in figure 1.

\vskip 0.3cm
\centerline{\epsfig{file=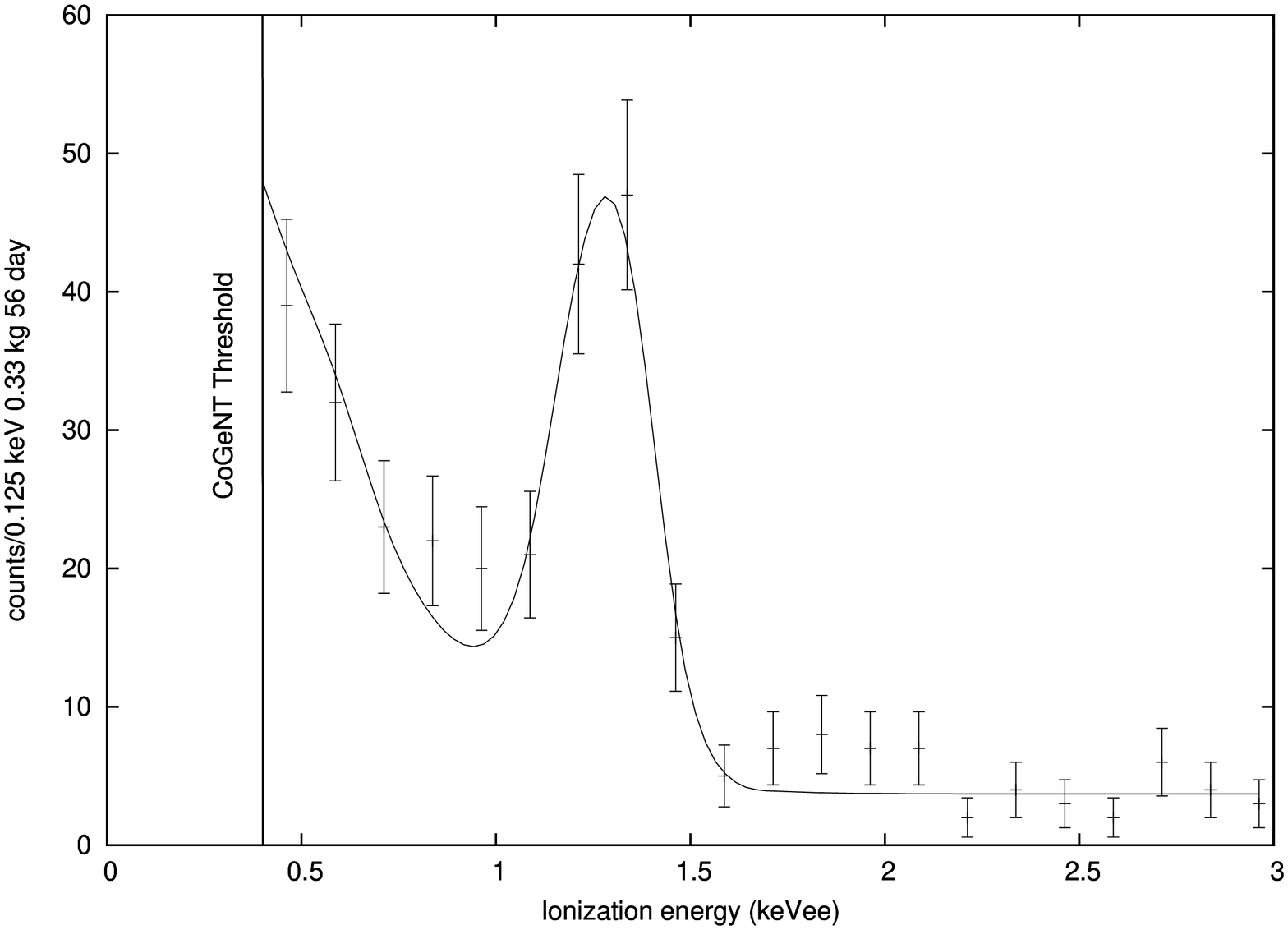,angle=270,width=13.0cm}}
\vskip 0.3cm
\noindent
Figure 1: Fit of the CoGeNT low energy spectrum for $m_{A'}/m_p = 20, \ \epsilon \sqrt{\xi_{A'}} = 6.2\times 10^{-10}$
and $v_{rot} = 254$ km/s. 
The background assumes two Gaussians to account for the 
$^{65}Zn$ (1.29 keV) and $^{68}Ge$ (1.1 keV) 
L-shell electron capture lines and a small constant rate.

\vskip 1.9cm

In order to work out the favoured region of the parameters $m_{A'}, \epsilon\sqrt{\xi_{A'}}$,
we define a $\chi^2$ quantity:
\footnote{
For the purposes of the fit, we analytically continue the mass number,
$A'$, to non-integer values,
with $Z' = A'/2$. Since the realistic case will involve a spectrum of
elements, the effective mass can be non-integer.} 
\begin{eqnarray}
\chi^2 (\epsilon\sqrt{\xi_{A'}}, m_{A'}) = \sum \left(
{\overline{dR}_i \over dE_R^m} - data_i \right)^2/\sigma^2_i
\end{eqnarray}
where ${\overline{dR}_i \over dE_R^m}$ is the differential rate averaged over the 
binned energy. The statistical errors are given by $\sigma_i = \sqrt{data_i}$.
We have minimized $\chi^2$ over a) the systematic uncertainties in quenching factor, which we take
as $q_{Ge} = 0.21 \pm 0.04$ and b) with respect
to the parameters of the background model describing the amplitudes of the $^{65}Zn$ (1.29 keV) and $^{68}Ge$ (1.1 keV) 
L-shell electron capture lines and adjusting also the constant background component.  
Note that no background exponential is assumed (or needed) to fit the data.
In the case of the DAMA analysis, the
annual modulation amplitude is used instead of the absolute rate, taken over twelve $0.5$ keVee 
energy bins
between 2-8 keVee. We have have minimized $\chi^2$ over the systematic uncertainties 
in the resolution and in the
Na and I quenching factors
which we take as $q_{Na} = 0.30 \pm 0.06, \ q_I = 0.09 \pm 0.02$. 
[See ref.\cite{collarnew} for a recent discussion
about quenching factor systematic uncertainties]. 
The best fit for CoGeNT has $\chi^2_{min} \simeq 12.5$ for 16 d.o.f  while the
best fit for DAMA has $\chi^2_{min} \simeq 8.5$ for 10 d.o.f.
An example near the DAMA best fit is shown in figure 2.

\vskip 0.2cm

\centerline{\epsfig{file=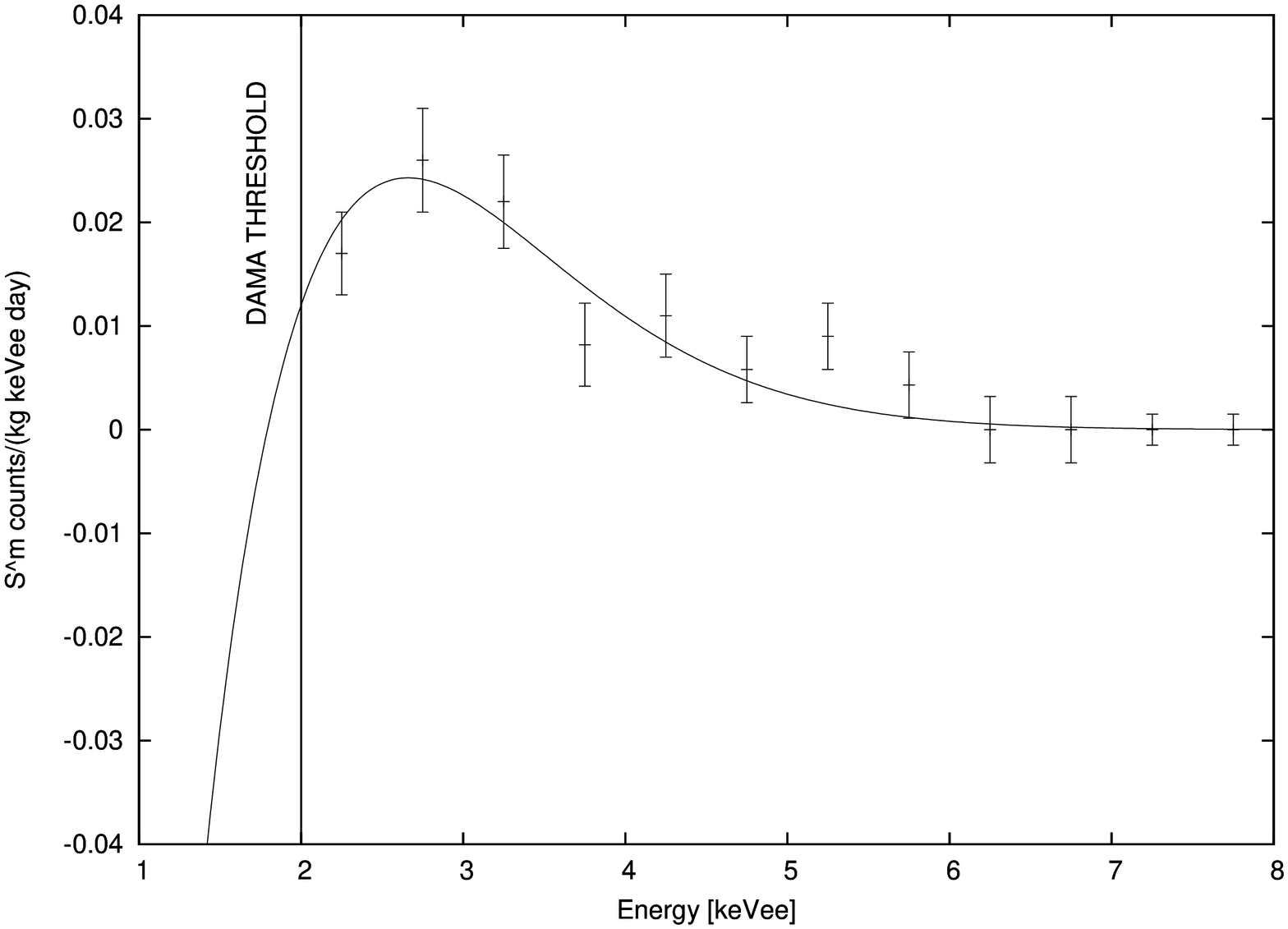,angle=270,width=12.0cm}}
\vskip 0.3cm

\noindent
Figure 2: Annual modulation amplitude versus measured recoil energy for the 
DAMA experiments. The solid line is the fit of the mirror dark matter candidate with
$m_{A'}/m_p = 20, \ \epsilon\sqrt{\xi_{A'}}  = 7.4\times 10^{-10}$ and $v_{rot} = 254$ km/s.
\vskip 0.4cm

Favoured regions in the 
$\epsilon\sqrt{\xi_{A'}}, m_{A'}$ plane can be obtained via evaluating contours corresponding to
$\chi^2 (\epsilon\sqrt{\xi_{A'}}, m_{A'}) = \chi^2_{min} + 9$ (roughly $99\%$ allowed C.L. region). 
We present the results in figure 3. 
As the figure demonstrates, CoGeNT and DAMA favoured regions are significantly overlapping 
which signifies that the mirror dark matter candidate provides a simultaneous explanation
of both the DAMA and CoGeNT signals.
Also shown in the figure is $95\%$ exclusion limits from CDMS/Si\cite{cdmssi},
CDMS/Ge\cite{cdmsge} and XENON100\cite{xenon100} experiments. In evaluating these exclusion limits, we have conservatively
taken the energy thresholds of these experiments to be $20\%$ higher than the advertised values, to
allow for systematic uncertainties in energy calibration and quenching factor
\footnote{Within the mirror dark matter framework the higher threshold experiments such as CDMS/Ge
and XENON100 have an important role in probing the heavier $Fe'$ component\cite{mm3}}.

\vskip 0.3cm

\centerline{\epsfig{file=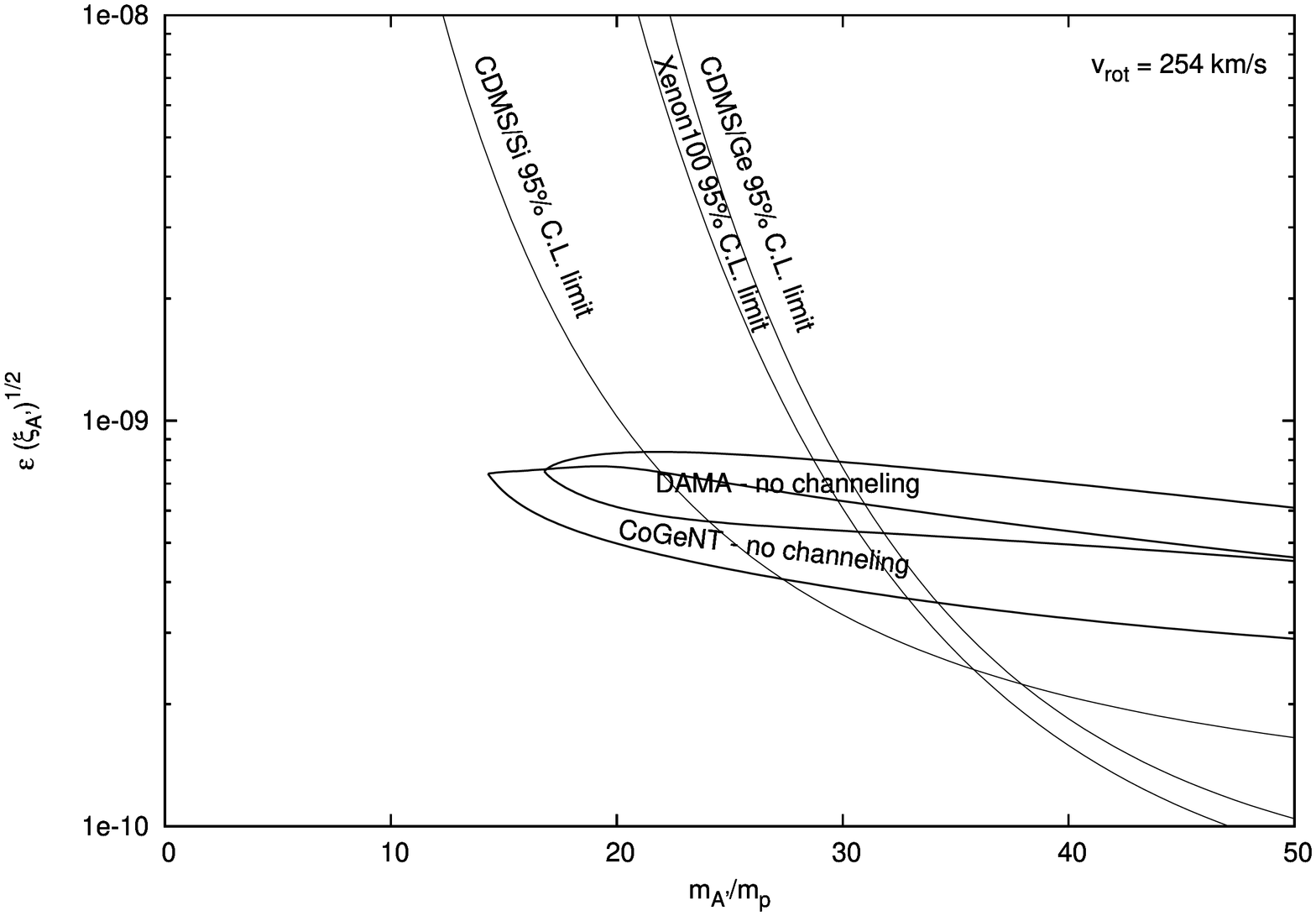,angle=270,width=12.3cm}}
\vskip 0.3cm

\noindent
Figure 3: CoGeNT and DAMA $99\%$ allowed C.L. regions of parameter space in the $\epsilon\sqrt{\xi_{A'}}, m_{A'}$ plane
for $v_{rot} = 254$ km/s. Negligible channeling is assumed.
Also shown is the $95\%$ exclusion region from the CDMS/Si,
CDMS/Ge and XENON100 experiments. [The region excluded is the right of the exclusion
limits]. 
\vskip 1.0cm

We have also performed a global analysis of the DAMA and CoGeNT data. Fixing $v_{rot} = 254$ km/s we have
evaluated $\chi^2$ for the combined DAMA + CoGeNT data, taking into account the systematic uncertainties
in quenching factor. We find
$\chi^2_{min} = 24.4$ for $28$ d.o.f. at $m_{A'}/m_p \simeq 24, \ \epsilon \sqrt{\xi_{A'}} = 
6.4 \times 10^{-10}$.
Evaluating the contours, $\chi^2 = \chi^2_{min} + 9$ we can determine the favoured 
region of parameter space in which the combination of DAMA and CoGeNT data are explained.
The result in shown in figure 4.

\vskip 0.2cm

\centerline{\epsfig{file=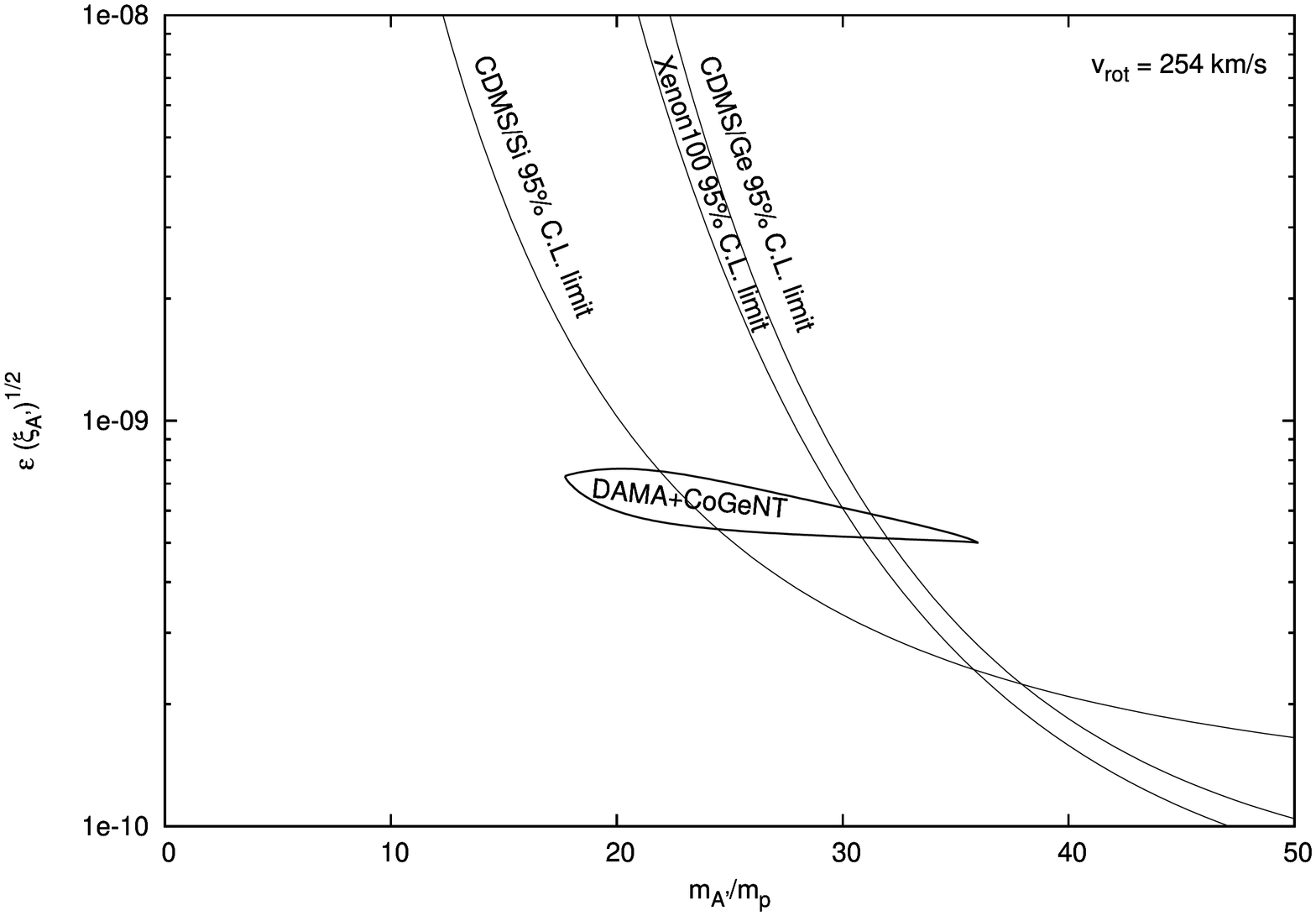,angle=270,width=13.0cm}}
\vskip 0.3cm

\noindent
Figure 4: Allowed regoin at 99\% C.L. in which the DAMA and CoGeNT signals are simultaneously explained, 
for $v_{rot} = 254$ km/s.  Negligible channeling is assumed.
Also shown is the $95\%$ exclusion region from the CDMS/Si,
CDMS/Ge and XENON100 experiments. [The region excluded is the right of the exclusion
limits]. 
\vskip 1.0cm

In conclusion, the recent low threshold CoGeNT Germanium experiment 
is the first to sensitively cover the recoil energy region in which the DAMA
experiments have obtained their annual modulation signal. Interestingly, CoGeNT
has found an unexplained rise in event
rate at low recoil energies - which is a tentative dark matter signal. 
We have shown that this rise in event rate is fully consistent
with the expectations from the mirror dark matter explanation of the DAMA experiment.
This result adds weight to the mirror dark matter interpretation of the direct detection experiments
which will be further testing by on-going experiments in the near future.

\vskip 1cm
\noindent
{\large Acknowledgments}

\vskip 0.2cm
\noindent
This work was supported by the Australian Research Council.

\end{document}